\documentclass[twocolumn,aps,showpacs]{revtex4}
\usepackage{amssymb}
\usepackage{amsmath,bm}
\usepackage{graphicx}
\usepackage[normalem]{ulem}

\usepackage[usenames, dvipsnames]{xcolor}
\usepackage{lipsum}

\renewcommand{\sout}{\bgroup \color{red} \ULdepth=-.5ex \ULset}

\begin{document}
\title{Nuclear dipole polarizability from mean-field modeling constrained by \\ chiral effective field theory}
\author{Zhen Zhang\footnote{zhenzhang$@$comp.tamu.edu}}
\author{Yeunhwan Lim\footnote{ylim@tamu.edu}}
\author{Jeremy W. Holt\footnote{holt@physics.tamu.edu}}
\author{Che Ming Ko\footnote{ko$@$comp.tamu.edu}}
\affiliation{Cyclotron Institute and Department of Physics and Astronomy, Texas A$\&$M University, College Station, Texas 77843, USA}
\date{\today}

\begin{abstract}
We construct a new Skyrme interaction Sk$\chi$m$^*$ by fitting the equation of state and nucleon effective masses in asymmetric nuclear matter from chiral two- and three-body forces as well as the binding energies of finite nuclei. Employing this interaction to study the electric dipole polarizabilities of $^{48}$Ca, $^{68}$Ni, $^{120}$Sn, and $^{208}$Pb in the random-phase approximation, we find that the theoretical predictions are in good agreement with experimentally measured values without additional fine tuning of the Skyrme interaction, thus confirming the usefulness of the new Skyrme interaction in studying the properties of nuclei. We further use this interaction to study the neutron skin thicknesses of $^{48}$Ca and $^{208}$Pb, and they are found to be consistent with the experimental data.
\end{abstract}

\maketitle

\section{Introduction}

The nuclear electric dipole polarizability $\alpha_{\mathrm{D}}$, which is proportional to the inverse energy-weighted sum of the dipole response of a nucleus to an external electric field, has recently attracted much attention due to its strong correlation with neutron skin thicknesses, the density dependence of the nuclear
symmetry energy~\cite{Rei10,Pie12, Zha14,Zha15, Pie14,Col14,Lat14,Roc13, Roc15}, and the properties of neutron stars \cite{Horowitz01,Fattoyev12,Erler13}. The electric dipole response is dominated by the nuclear isovector giant dipole resonance (IVGDR), which is the oldest known nuclear collective excitation and has been extensively studied both theoretically
and experimentally~\cite{Liu75, Speth81, Woude87, Chak87}. It is well known from theoretical studies based on various models that the properties of the IVGDR, in which neutrons oscillate against protons in a nucleus, is affected by the density dependence of the nuclear symmetry energy~\cite{Mey82,Kri84,Lip89,Dan03,Tri08}.
%\cite{Horowitz01,Fattoyev12,Erler13}.

Experimentally, the electric dipole strength distributions in $^{48}$Ca~\cite{Bir16}, $^{120}\mathrm{Sn}$~\cite{Has15}, and $^{208}\mathrm{Pb}$~\cite{Tam11} have recently been  measured  accurately at the Research Center for Nuclear Physics (RCNP) from proton inelastic scattering experiments at forward angles, while that of $^{68}$Ni has been investigated at GSI using Coulomb excitations in inverse kinematics and measuring the invariant mass in the one- and two-neutron decay channels~\cite{Ros13}. Recently, ab initio calculations based on chiral effective field theory interactions have been successful at describing $\alpha_{\mathrm{D}}$ in medium-mass nuclei~\cite{Hag16,Mio16}, but the description of the dipole response of heavy nuclei remains a challenge. Therefore, nuclear energy density functionals are still the most widely-used approach to explore the nuclear equation of state (EOS) from the electric dipole response.

In the present work, we construct a new Skyrme interaction Sk$\chi$m$^*$ by fitting the equation of state and nucleon effective masses in asymmetric nuclear matter predicted by
chiral two- and three-body forces as well as the binding energies of finite nuclei. We then employ it to study the dipole response of $^{48}$Ca, $^{68}$Ni, $^{120}$Sn, and $^{208}$Pb within the random-phase approximation (RPA). The calculated electric dipole polarizabilities are in good agreement with existing experimental values. In addition, the neutron skin thickness in $^{48}$Ca and $^{208}$Pb is discussed.

The paper is organized as follows. In Section \ref{sec2}, we explain in detail the new Skyrme interaction Sk$\chi$m$^*$ and how it is obtained. In Section III, we describe the random-phase approximation  employed in our calculations, and in Section IV we present results and discussions for the isovector dipole response, electric dipole polarizability, and neutron skin thicknesses of $^{48}$Ca and $^{208}$Pb  from the new Skyrme mean-field model.  We end with a summary in Section V.

\section{A Skyrme interaction from chiral effective field theory}
\label{sec2}

Recently, extended Skyrme interactions have been constructed in Ref.~\cite{Lim17} by fitting to the asymmetric nuclear matter equation of state from chiral two- and three-nucleon forces \cite{Wellenhofer15,Wellenhofer16}. In contrast to previous works \cite{Brown14,Rrapaj16}
imposing constraints on mean-field models from low-density neutron matter, the
full density dependence of the asymmetric matter equation of state was used in Ref.\ \cite{Lim17} for the $\chi^2$ minimization function. In these calculations, the chiral
two-body force was treated at next-to-next-to-next-to-leading order (N3LO) in the chiral power counting, while the three-body force was treated at N2LO. Three different choices of the momentum-space cutoff were considered \{$\Lambda= 414,~450 ~\mathrm{and} ~500~\mathrm{MeV}$\}, and the unknown low-energy constants associated with short-distance dynamics were fitted \cite{Entem03,Coraggio07,Coraggio13,Coraggio14} in each case to nucleon-nucleon scattering phase shifts and deuteron properties (in the case of the
nucleon-nucleon interaction) as well as the binding energy and beta-decay lifetime of $^3$H (in the case of the three-nucleon force). To fix the gradient contributions to the energy density functional, the extended Skyrme interactions were also fitted to reproduce
the binding energies of 7 doubly-closed-shell nuclei: $^{16}\mathrm{O}$, $^{40}\mathrm{Ca}$,
$^{48}\mathrm{Ca}$, $^{56}\mathrm{Ni}$, $^{100}\mathrm{Sn}$, $^{132}\mathrm{Sn}$, and
$^{208}\mathrm{Pb}$. The conventional Skyrme interaction~\cite{Cha97} was found to be inadequate to describe the density dependence of the equation of state from chiral effective field theory, leading the authors of Ref.\ \cite{Lim17} to include an additional density-dependent and momentum-independent term. The three Skyrme interactions (Sk$\chi414$, Sk$\chi450$ and Sk$\chi500$) reported in Ref.~\cite{Lim17} predict, however, too large (even larger than the nucleon bare mass $m$) isoscalar and isovector effective masses $m_{s,0}^{\ast}$ and $m_{v,0}^{\ast}$ at saturation density, which are given by~\cite{Cha97}
\begin{eqnarray}
\frac{1}{m_{s,0}^{\ast}} & =& \frac{1}{m}+\frac{1}{8}
[3t_1+t_2(5+4x_2)]\rho_0, \\
\frac{1}{m_{v,0}^{\ast}} & =& \frac{1}{m}+\frac{1}{4}
[t_1(2+x_1)+t_2(2+x_2)]\rho_0.
\end{eqnarray}
This fact makes Sk$\chi414$, Sk$\chi450$ and Sk$\chi500$ unsuitable for studying the nuclear giant dipole resonances. It is well known that the $m_{v,0}^{\ast}$ is related to the enhancement factor $\kappa$ in the Thomas-Reiche-Kuhn sum rule~\cite{Ring}, given by the relation $1+\kappa = m/m_{v,0}^{\ast}$, and thus affects the nuclear dipole response function. We note that in Skyrme functionals  the neutron and proton effective masses $m_n^*$ and  $m_p^*$ at total baryon number density $\rho$ are related to the isoscalar and isovector effective masses $m_{s}^*(\rho)$ and $m_{v}^*(\rho)$  by~\cite{Farine01}
\begin{eqnarray}\label{Eq:mn}
\frac{1}{m_n^*}&=&(1+\delta)\frac{1}{m_s^*}-\delta\frac{1}{m_v^*}, \\
\frac{1}{m_p^*}&=&(1-\delta)\frac{1}{m_s^*}+\delta\frac{1}{m_v^*}\label{Eq:mp}
\end{eqnarray}
where $\delta =(\rho_n-\rho_p)/(\rho_n+\rho_p)$ is the isospin asymmetry with $\rho_n~(\rho_p)$ being the neutron (proton) density.

In the present work, instead of directly fitting the three EOSs predicted by chiral effective field theory as in Ref~\cite{Lim17}, we treat them as theoretical uncertainties on the equation of state and construct a Skyrme interaction by fitting the central values. As shown later, taking into account the theoretical uncertainties makes it possible to use the conventional Skyrme interaction to reproduce the predicted EOS from chiral effective field theory.  To improve the description of nucleon effective masses, we further include in our fit the isoscalar and isovector effective masses $m_{s,0}^{\ast}/m = 0.82\pm0.08$ and $m_{v,0}^{\ast}/m = 0.69\pm 0.02$ at saturation density that are extracted from the single-particle energies of protons and neutrons in asymmetric nuclear matter, computed recently in Refs.\ \cite{holt13prc,holt16prc} from chiral two- and three-body forces.  From the momentum- and energy-dependent nucleon single-particle energy $e(p) = \frac{p^2}{2m}+\mathrm{Re} \Sigma( p, e(p) )$, where $p$ is the nucleon momentum and the self-energy $\Sigma$ is computed at second order in perturbation theory, the nucleon effective mass $m^*$ can be extracted by the definition  
\begin{equation}
\frac{p}{m^*}=\frac{d e}{dp}.
\end{equation}
The $m_{s,0}^{\ast}$ and $m_{v,0}^{\ast}$  then can be obtained by invoking Eqs.~(\ref{Eq:mn}) and (\ref{Eq:mp}). Since the effective masses in the Skyrme-Hartree-Fock model are momentum-independent, we average the momentum-dependent effective masses from chrial effective theory over a momentum range around the Fermi momentum, which accounts for most of the theoretical uncertainties.  It is interesting to see that the $m^*_{v,0}$ from chiral effective field theory is consistent with that from an analysis of dipole resonances in $^{208}$Pb based on conventional Skyrme interactions~\cite{Klu09}.

Our new Skyrme interaction therefore has the usual form:
\begin{eqnarray}
v(\bm{r}_1,\bm{r}_2)&=&t_0(1+x_0P_{\sigma})\delta(\bm{r}_1-\bm{r}_2)   \notag \\
& &+\frac{1}{2}t_1(1+x_1P_{\sigma})[\bm{k}'^2\delta(\bm{r}_1-\bm{r}_2)+\mathrm{c.c.}] \notag \\
& &+t_2(1+x_2P_{\sigma})\bm{k}'\cdot\delta(\bm{r}_1-\bm{r}_2)\bm{k} \notag \\
& &+\frac{1}{6}t_3(1+x_3P_{\sigma})\rho ^{\alpha}\left(\frac{\bm{r}_1+\bm{r}_2}{2}\right)\delta(\bm{r}_1-\bm{r}_2)\notag\\
& &+iW_0(\bm{\sigma}_1+\bm{\sigma}_2)\cdot[\bm{k}'\times\delta(\bm{r}_1-\bm{r}_2)\bm{k}],
\label{Eq:Sky}
\end{eqnarray}
where $\bm{\sigma}_i$ is the Pauli spin operator, $P_{\sigma}=(1+\bm{\sigma}_1\cdot\bm{\sigma}_2)/2$ is the spin-exchange operator, $\bm{k}=-i(\bm{ \nabla}_1-\bm{\nabla}_2)/2$ is the relative momentum operator, and $\bm{k}^{\prime}$ is the conjugate operator of $\bm{k}$ acting on the left.

Following Refs.~\cite{LWC10, Kor10}, we  express the 9 parameters $t_0- t_3$, $x_0- x_3$ and $\alpha$ of the  Skyrme interaction in terms of 9 macroscopic quantities: $m_{s,0}^{\ast}$, $m_{v,0}^{\ast}$,  the nuclear matter saturation density $\rho_0$, the energy per particle of symmetric nuclear matter $E_0(\rho_0)$, the incompressibility $K_0$, the gradient coefficient $G_S$, the symmetry-gradient coefficient $G_{V}$, and the magnitude $E_{\mathrm{sym}}(\rho_0)$ and density slope $L$ of the nuclear symmetry energy at $\rho_0$~\cite{LWC10, Kor10}. Here the $G_S$ and $G_V$ are defined by expressing the momentum-dependent or finite-range terms of the Skyrme energy density functional $\mathcal{H}_{\mathrm{fin}}$ as~\cite{LWC10}
\begin{equation}
\mathcal{H}_{\mathrm{fin}}= \frac{G_S}{2}(\nabla \rho)^2
-\frac{G_V}{2}(\nabla \rho_n-\nabla\rho_p)^2
\end{equation}
with
\begin{eqnarray}
G_S&=&\frac{9}{32}t_1-\frac{1}{32}t_2(4x_2+5), \\
G_V&=&\frac{3}{32}t_1(2x_1+1)+\frac{1}{32}t_2(2x_2+1).
\end{eqnarray}
Consequently, the Skyrme mean-field models have the following 10 parameters:
\begin{eqnarray}
\bm{p} & =& \lbrace  \rho_0,~E_0(\rho_0),~K_0,~E_{\mathrm{sym}}(\rho_0),~L, \notag \\
& &~ G_S,~G_V,~W_0,~m_{s,0}^{\ast},~m_{v,0}^{\ast}\rbrace.
\end{eqnarray}
These parameters have clear physical meanings and their ranges are empirically known, which increases significantly the efficiency of the optimization algorithm. The parameters $G_S$, $G_V$ and $W_0$, which are not constrained by the nuclear equation of state, have been studied \cite{Kai11, Kai12, Holt16} consistently in chiral effective field theory via the density matrix expansion in the Hartree-Fock approximation. However, higher-order terms in the expansion are needed to produce reliable constraints on the associated parameters in Skyrme functionals. In the present work we constrain the values of $G_S$, $G_V$ and $W_0$ by fitting to the binding energies of double-closed-shell nuclei as in Ref.~\cite{Lim17}.

As in the usual fitting procedure, we optimize the parameters $\bm{p}(p_1,p_2,\cdots, p_N)$
by minimizing the weighted sum of the squared deviations from the data as
\begin{eqnarray}
\chi^2 &=& \sum_{i=1}^{n_E}\left(\frac{E^{\mathrm{Sky}}_i(\rho_i,\delta_i)-E^{\mathrm{EFT}}_i(\rho_i,\delta_i)}{\sigma_E}\right)^2 \notag\\
& & +\sum_{i=1}^{n_B}\left(\frac{B^{\mathrm{Sky}}_i-B^{\mathrm{exp}}_i}{\sigma_B}\right)^2+\left( \frac{m_{s,0}^*/m-0.82}{0.08}\right)^2 \notag\\
& &+\left(\frac{m_{v,0}^*/m-0.69}{0.02} \right)^2,
\end{eqnarray}
where $E_i$ is the binding energy per nucleon in nuclear matter with the nucleon density $\rho_i=\{0.01,~0.04,~0.08,~0.16,~0.20,~0.24,~0.28,~0.32\}~\mathrm{fm}^{-3}$ and the isospin asymmetry $\delta_i^2=\{0,~0.25,~0.5,~0.75,~1\}$; $B_i$ is the total binding energy of the $i$th nucleus; `Sky' and `EFT' indicate predictions of the Skyrme-Hartree-Fock model and chiral effective field theory, while `exp' indicates the experimental data. The adopted error $\sigma_B$ is set to be $1$ MeV, while $\sigma_E$ is taken to be max($0.25, \sigma_{\mathrm{EFT}}(\rho)$)~MeV where $\sigma_{\mathrm{EFT}}(\rho)$ is the theoretical error of the EOS due to the variation of the momentum cutoff used in chiral effective field theory.  Since the predictions of chiral effective  field theory are very precise at low densities (e.g., the error is $0.02$ MeV in neutron matter at $\rho=0.01$ fm$^{-3}$), a minimum error of $0.25$ MeV for the EOS is introduced to take account of the deficiency of the Skyrme parametrization. The optimization is carried out by employing the POUNDERS algorithm in the distribution of PETSc/Tao~\cite{Tao}, which has been successfully used previously to calibrate nuclear energy density functionals~\cite{Kor10,Kor12,Kor14}.

Once the optimized parameter set $\bm{p}_0$ is obtained, the $\chi^2$ near $\bm{p}_0$ can be approximately written as
\begin{equation}
\chi^2(\bm{p})\approx\chi^2(\bm{p}_0)+\frac{1}{2}
\sum_{i,j}^{N}(p_i-p_{0i})\mathcal{M}_{ij}(p_j-p_{0j}),
\end{equation}
with $\mathcal{M}_{ij} = \left.\frac{\partial^2\chi^2}{\partial{p_i}\partial{p_j}} \right\vert_{\bm{p}_0}$. The
errors of parameters then can be estimated as
\begin{equation}
e_i=\sqrt{(\mathcal{M}^{-1})_{ii}} \, ,
\end{equation}
which is the allowed variation of the parameter $p_i$ within the ellipsoid of $\chi^2 -\chi^2(\bm{p}_0) \leq 1$.  Using the covariance matrix $\Sigma = \mathcal{M}^{-1}$, the correlation matrix can be estimated as
\begin{equation}
\mathcal{C}_{ij}=\frac{\Sigma_{ij}}{\sqrt{\Sigma_{ii}\Sigma_{jj}}},
\end{equation}
where $\mathcal{C}_{ij}$ is the correlation coefficient between parameters $p_i$ and $p_j$. The extrapolation error of any observables and the correlation coefficient between them can also be determined (see e.g., \cite{Dob14,Pie15}).

\begin{table}[h]
\caption{Parameters of the Skyrme interaction Sk$\chi m^*$ and their errors $\sigma$: lines 1-9 show the chosen 9 macroscopic quantities used to express the Skyrme parameters (see text for details); lines 10-19 show the Skyrme parameters.}
\label{Tab:Sky}
\begin{tabular}{lcc}
\hline \hline
Quantity & Sk$\chi m^*$ & $\sigma$ \\
\hline 
$\rho_0~(\mathrm{fm}^{-3})$                       &       0.1651     &       0.0025    \\
$E_0~(\mathrm{MeV}) $                             &     -16.07       &       0.06      \\
$K_0~(\mathrm{MeV})$                              &     230.4        &       6.6       \\
$E_{\mathrm{sym}}(\rho_0)~(\mathrm{MeV})$         &      30.94       &       0.46  \\
$L~(\mathrm{MeV})$                                &      45.6        &       2.5    \\
$G_S(\mathrm{MeV}\cdot \mathrm{fm}^{5})$          &     141.5        &       7.3   \\
$G_V(\mathrm{MeV}\cdot \mathrm{fm}^{5})$          &     -70.5        &      32.6   \\
$m_{s,0}^{\ast}/m$                                &       0.750      &         0.040  \\
$m_{v,0}^{\ast}/m$                                &       0.694      &       0.020 \\
\hline
$W_0(\mathrm{MeV}\cdot \mathrm{fm}^{5})$          &     119.8      &       4.9  \\
$t_0(\mathrm{MeV}\cdot \mathrm{fm}^{3})$          &   -2260.7        &     365.8  \\
$t_1(\mathrm{MeV}\cdot \mathrm{fm}^{5})$          &     433.189      &      30.661  \\
$t_2(\mathrm{MeV}\cdot \mathrm{fm}^{5})$          &     274.553      &     173.611  \\
$t_3(\mathrm{MeV}\cdot \mathrm{fm}^{3+3\alpha})$  &   12984.4        &    1762.4  \\
$x_0$                                             &       0.327488   &       0.110201  \\
$x_1$                                             &      -1.088968   &       0.303910  \\
$x_2$                                             &      -1.822404   &       0.366035  \\
$x_3$                                             &       0.442900   &       0.215646  \\
$\alpha$                                          &       0.198029   &       0.054443  \\

\hline\hline
\end{tabular}
\end{table}

\begin{figure}[h]
\includegraphics[width=8.7cm]{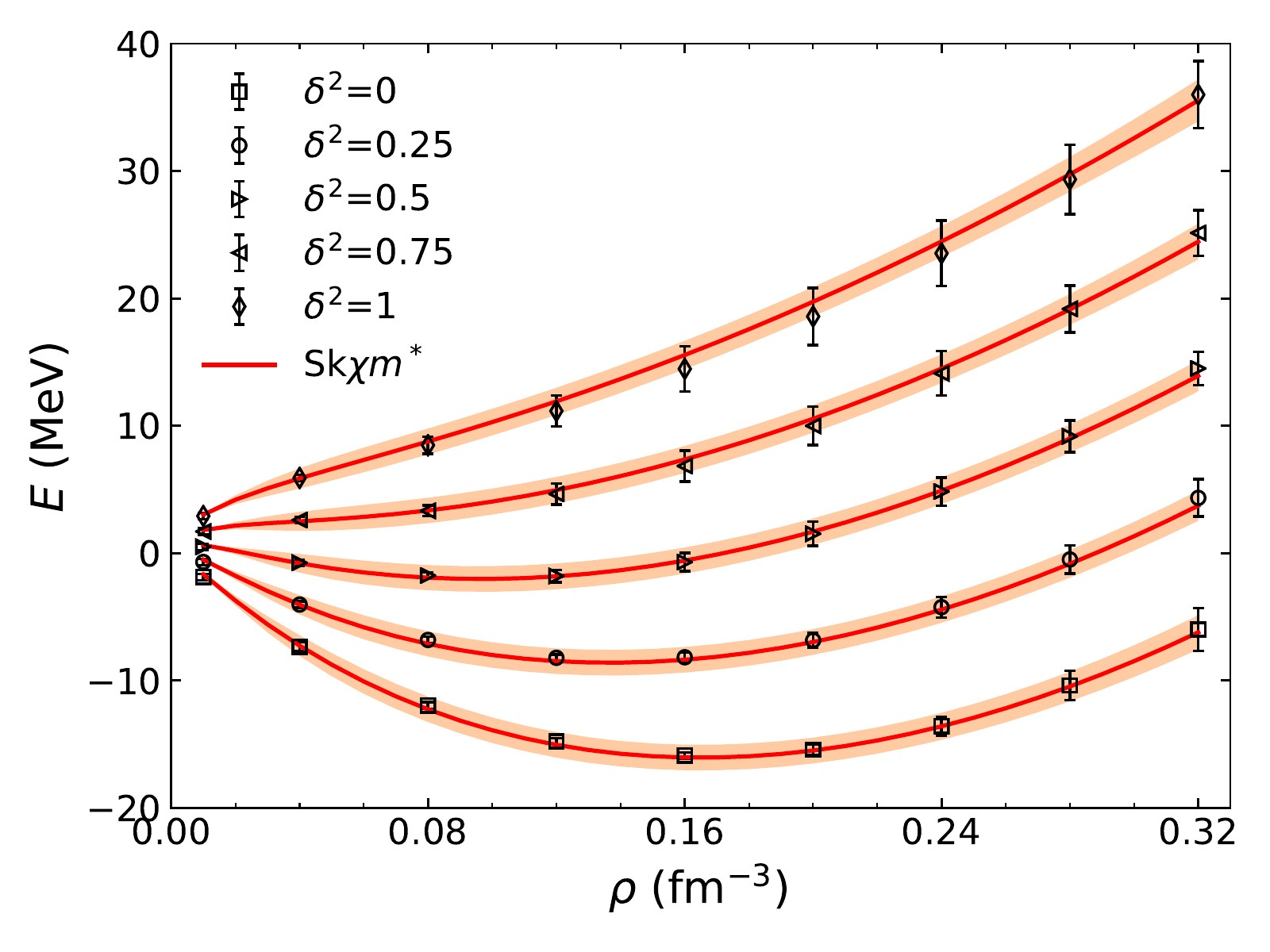}
\caption{(Color online.) EOS of asymmetric nuclear matter predicted by Sk$\chi m^*$. For comparison, predictions from the chiral effective field theory are also shown.}\label{Fig:EOS}
\end{figure}

The obtained parameter set $\bm{p}_0$, named Sk$\chi m^*$, including its uncertainties, together with the corresponding Skyrme parameters are listed in Tab.~\ref{Tab:Sky}. It is seen that the parameters $\rho_0,~E_0(\rho_0),~K_0,~E_{\mathrm{sym}}(\rho_0)$ and $L$, which characterize the density- and isospin-dependence of the nuclear EOS, are well constrained by the predicted EOS from chiral effective field theory.  If we had removed from the fit all data points shown in Fig.~\ref{Fig:EOS} for the equation of state of asymmetric matter, the uncertainty in $L$ would be roughly 60 MeV.  Also, the isoscalar gradient coefficient $G_S=141.5\pm 7.3~\mathrm{MeV}\cdot \mathrm{fm}^{5}$ and the spin-orbit coupling strength $W_0 = 119.8\pm4.9(\mathrm{MeV}\cdot \mathrm{fm}^{5})$  are consistent with the values at sub-saturation densities of about $0.03$~fm$^{-3}$ in the energy density functional derived from chiral two- and three-body forces~\cite{Holt16}. We show in Fig.~\ref{Fig:EOS} the EOSs of asymmetric nuclear matter predicted by the Sk$\chi m^*$ interaction together with predictions from chiral effective field theory. One can see that the new interaction can well reproduce the EOSs from chiral effective field theory calculations. In Tab.~\ref{Tab:EB}, we list the binding energies and charge radii of  $^{16}\mathrm{O}$, $^{40}\mathrm{Ca}$,$^{48}\mathrm{Ca}$, $^{56}\mathrm{Ni}$, $^{100}\mathrm{Sn}$, $^{132}\mathrm{Sn}$, and $^{208}\mathrm{Pb}$ obtained by Hartree-Fock calculations using Sk$\chi m^*$, together with their experimental values. It is seen that our results are overall in very good agreement with the experimental data for the  7 doubly-closed-shell nuclei.

\begin{table}[h]
\caption{Comparisons of binding energies $B$ (in MeV) and charge radii $r_c$ (in fm) of $^{208}$Pb, $^{132}$Sn, $^{100}$Sn, $^{56}$Ni, $^{48}$Ca, $^{40}$Ca and $^{16}$O from Hartree-Fock calculations with the Sk$\chi m^{\ast}$ interaction to experimental data.}
\label{Tab:EB}
\begin{tabular}{lcccc}
\hline \hline
&$B^{\mathrm{exp}}$ & $B$ & $r_c^{\mathrm{exp}}$ & $r_c$\\
\hline
$^{208}$Pb & -1636.44   &  -1635.71&5.50  &5.46 \\
$^{132}$Sn & -1102.90   &  -1103.47&4.71 &4.68\\
$^{100}$Sn & -825.78   &  -826.77&- &4.45\\
$^{56}$Ni  & -483.99   &  -482.83&- &3.74\\
$^{48}$Ca  & -415.99   &  -416.55&3.48 &3.47\\
$^{40}$Ca  & -342.05  &  -342.08& 3.48&3.46\\
$^{16}$O   & -127.62   &  -127.24&2.70 &2.74 \\
\hline
\hline
\end{tabular}
\end{table}

\begin{figure}[h]
\includegraphics[width=\linewidth]{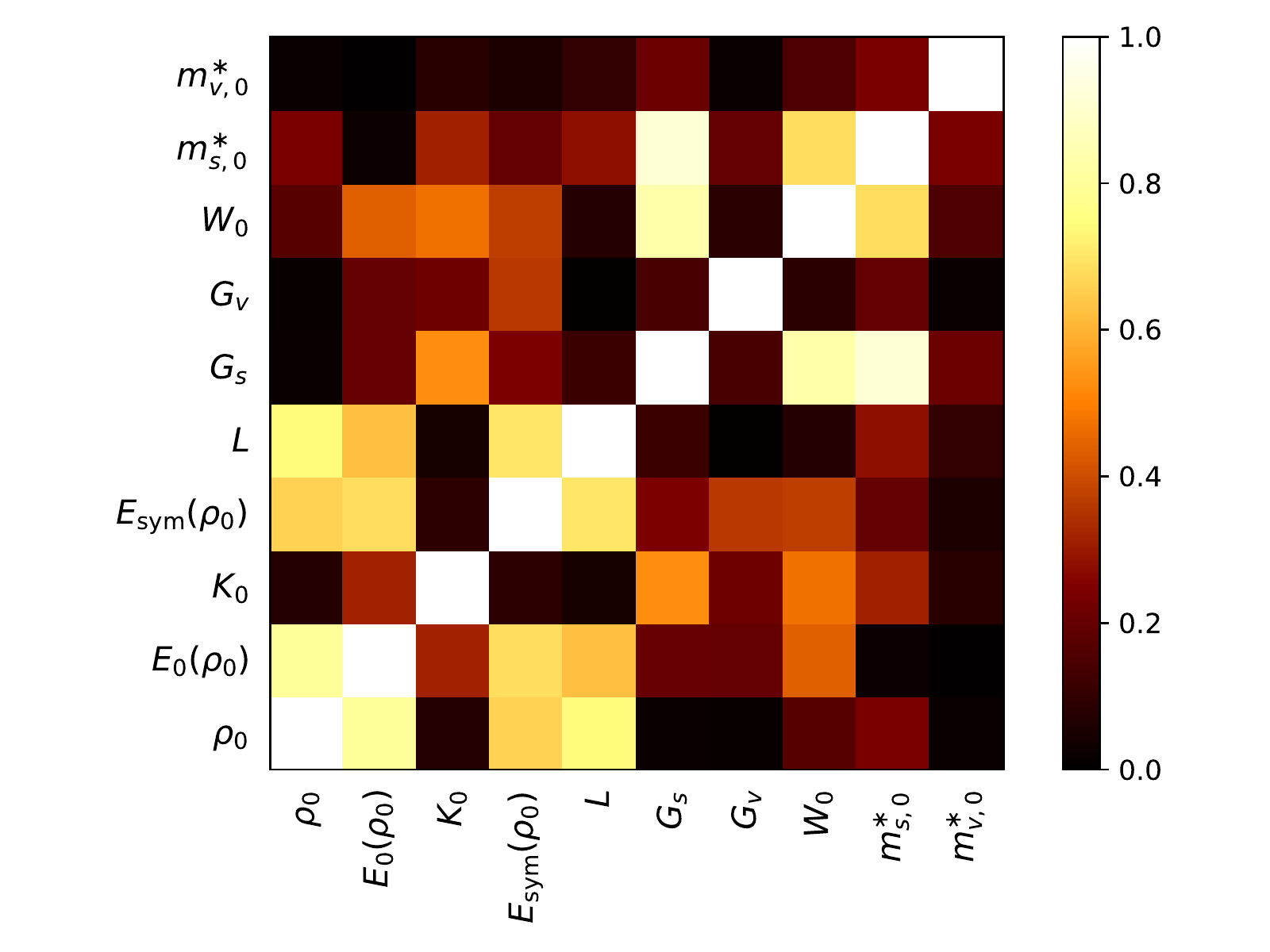}
\caption{(Color online.) Absolute values of correlation coefficients among the 10 parameters $\rho_0$, $E_0(\rho_0)$, $K_0$, $E_{\mathrm{sym}}(\rho_0)$, $L$, $K_{\mathrm{sym}}$, $G_S$, $G_V$, $W_0 $, $m_{s,0}^*$ and $m_{v,0}^{\ast}$.}
\label{Fig:Cor}
\end{figure}

Fig.~\ref{Fig:Cor} shows the absolute values of correlation coefficients calculated among the 10 parameters $\rho_0$, $E_0(\rho_0)$, $K_0$, $E_{\mathrm{sym}}(\rho_0)$, $L$, $G_S$, $G_V$, $W_0 $, $m_{s,0}^{\ast}$ and $m_{v,0}^{\ast}$. We find that $m_{v,0}^{\ast}$ is almost independent of all other parameters, which reflects the fact that it is poorly constrained by the data from the EOSs and nuclear binding energies used in our fit. Except $m_{v,0}^{\ast}$, most parameters are correlated with each other, but generally the correlations are  weak with the absolute value of the Pearson coefficient $R$ less than 0.8. The spin-orbit coupling strength $W_0$ is strongly correlated with both the isoscalar effective mass $m_{s,0}$ and the isoscalar gradient coefficient $G_S$, which are related to the momentum-dependent terms in the Skyrme interaction. 

\section{Isovector dipole response  and random-phase approximation}

The dipole response of a nucleus can be calculated in the random-phase approximation using the dipole operator
\begin{equation}
\hat{F} = \frac{N}{A}\sum^Z_{i=1}r_iY_{\text{1M}}(\hat{r}_i)-\frac{Z}{A}\sum^N_{i=1}r_iY_{\text{1M}}(\hat{r}_i),
\end{equation}
where $r_i$ is the nucleon's radial coordinate and $Y_{\text{1M}}(\hat{r_i})$ is the corresponding spherical harmonic function. In the RPA method, the isovector dipole strength function is evaluated as
\begin{equation}
S(E)=\sum_{\nu}|\langle\nu\Vert\hat{F}\Vert\tilde{0}\rangle|^2\delta(E-E_{\nu}),
\end{equation}
where $|\tilde{0}\rangle $ is the RPA ground state, and  $|\nu \rangle$ is the RPA excited state  with $E_{\nu}$ being its energy. Defining the moments of the strength function as
\begin{equation}
m_k=\int dE E^kS(E)=\sum_{\nu}|\langle\nu\Vert\hat{F}\Vert\tilde{0}\rangle|^2E_{\nu}^k,\label{Eq:mk}
\end{equation}
the electric dipole polarizability $\alpha_{D}$ can then be calculated according to
\begin{equation}
\label{Eq:AlphaDSm1}
\alpha_{D}=\frac{8\pi}{9}e^2\int dEE^{-1}S(E)=\frac{8\pi}{9}e^2m_{-1}.
\end{equation}

\begin{table}[h]
\caption{Experimental data for the electric dipole polarizabilities $\alpha_{\mathrm{D}}$
(in units of fm$^3$) for $^{48}\mathrm{Ca}$~\cite{Bir16}, $^{68}\mathrm{Ni}$~\cite{Ros13}, $^{120}\mathrm{Sn}$~\cite{Has15}, and $^{208}\mathrm{Pb}$~\cite{Tam11} together with the results of RPA calculations using the Sk$\chi m^{\ast}$ Skyrme parametrization .}
\label{Tab:Data}

\begin{tabular}{lcc}
\hline
\hline
                    & ~Expt.~    & ~Sk$\chi m^{\ast}$ \\
 \hline
$^{48}\mathrm{Ca}$  & 2.07(22)   & 2.27(5)          \\
$^{68}\mathrm{Ni}$  & 3.88(31)   & 4.06(7)	        \\
$^{120}\mathrm{Sn}$ & 8.59(36)   & 9.28(14)	        \\
$^{208}\mathrm{Pb}$ & 19.6(6)    & 19.87(29)        \\
\hline\hline
\end{tabular}
\end{table}

In the present work, we employ the Skyrme-RPA code by Colo \textit{et al.}~\cite{Col13} 
to calculate the electric dipole polarizabilities of $^{48}\mathrm{Ca}$, $^{68}\mathrm{Ni}$, $^{120}\mathrm{Sn}$, and $^{208}\mathrm{Pb}$.  As pointed out in Ref.~\cite{Roc15}, to compare the RPA results with experiment, the experimentally extracted values of
$\alpha_{\mathrm{D}}$ for $^{120}\mathrm{Sn}$~\cite{Has15} and $^{208}\mathrm{Pb}$~\cite{Tam11} need to be corrected by subtracting the contributions from  quasi-deuteron excitations~\cite{Lep81,Sch88}, while that of $^{68}\mathrm{Ni}$~\cite{Hag16} should be modified by including the corrections from the extrapolated low-energy and high-energy regions. The corrected experimental data are shown in Table~\ref{Tab:Data}.

\section{Results and discussions}

\begin{figure}[h]
\includegraphics[width=8.7cm]{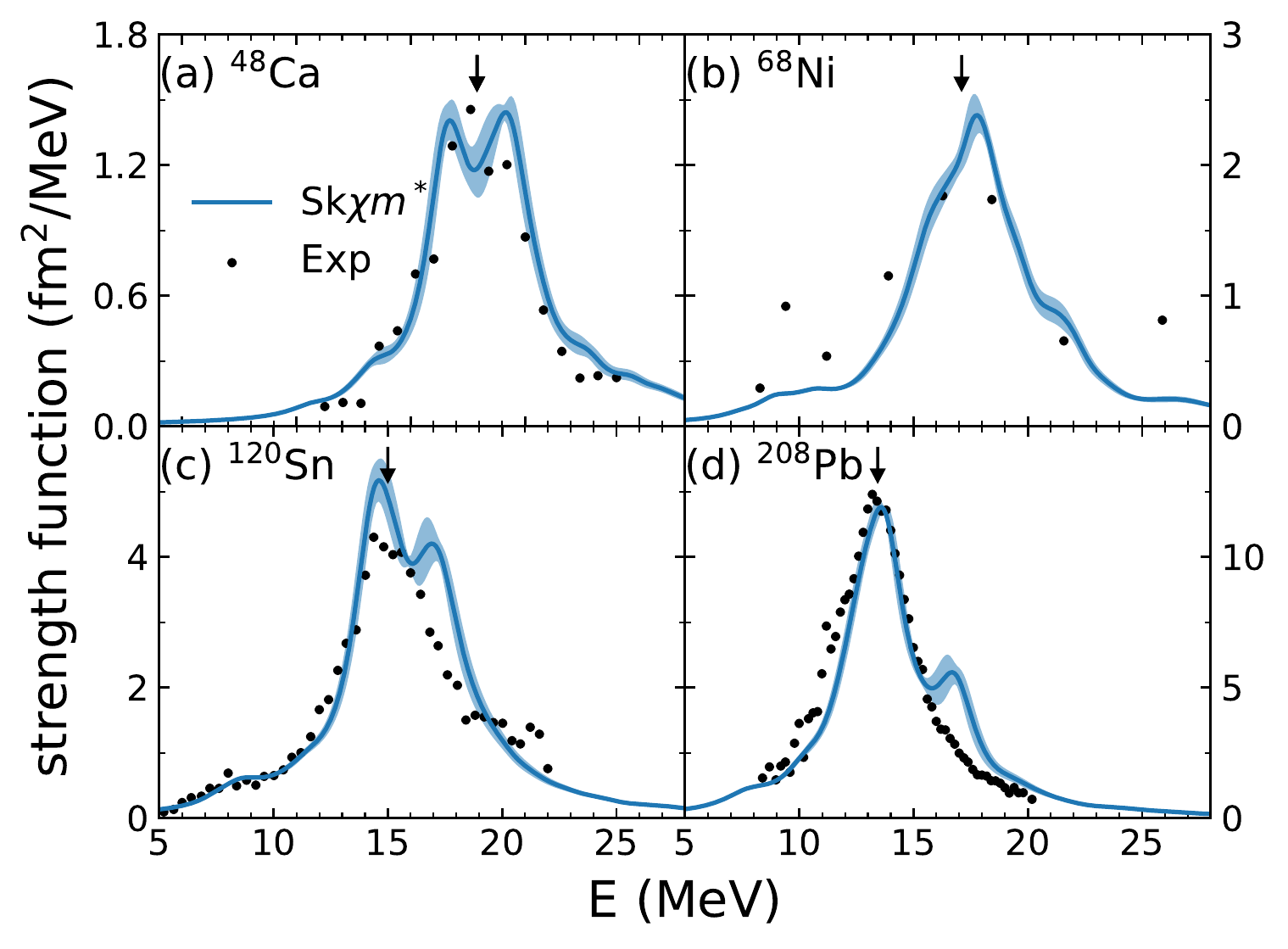}
\caption{(Color online.) Strength functions of the IVGDR in $^{48}\mathrm{Ca}$, $^{68}\mathrm{Ni}$, $^{120}\mathrm{Sn}$, and $^{208}\mathrm{Pb}$ obtained from
RPA calculations using Sk$\chi m^*$. For comparison, experimental data~\cite{Tam11,Ros13,Has15,Bir16} are shown as black solid circles.}
\label{Fig:Sgdr}
\end{figure}

In Fig.~\ref{Fig:Sgdr}, we show the isovector dipole transition strength functions for $^{48}\mathrm{Ca}$, $^{68}\mathrm{Ni}$, $^{120}\mathrm{Sn}$, and $^{208}$Pb
obtained from the RPA calculations using the Skyrme interaction Sk$\chi m^*$.  Here the curves are obtained by smearing each of the discrete RPA peaks by a Lorentzian function  of 2 MeV width, which is taken to reproduce the experimental width of the IVGDR.  For comparison, the measured strength functions for these nuclei are shown as black solid circles, and the experimental peak energies $E_x=18.9, ~17.1, ~15$ and $13.43~\mathrm{MeV}$~\cite{Tam11, Ros13, Bir16, Has15, Die88} of the IVGDR in $^{48}\mathrm{Ca}$, $^{68}\mathrm{Ni}$, $^{120}\mathrm{Sn}$, and $^{208}\mathrm{Pb}$ are indicated by the arrows. It is seen that the Sk$\chi m^*$ interaction gives  dipole strength functions which are consistent with experimental measurements, except the predicted second peak at higher energy that is not found in experimental data, especially for $^{120}$Sn.  Such an unphysical high-energy peak is expected to become a long high-energy tail if  more complex configurations beyond the $1p-1h$ states are included in the RPA calculation~\cite{Lyu16}. For $^{208}$Pb,  the predicted peak energy $E_x$ from Sk$\chi m^*$ at 13.6~$\mathrm{MeV}$ is in particularly good agreement with the experimental value. This is due to the $m_{v,0}^{\ast}=0.694m$, or equivalently $\kappa =0.44$, predicted by chiral effective field theory, which is roughly consistent with the value $\kappa =0.4$ determined by fitting the IVGDR peak energy in $^{208}$Pb using conventional Skyrme interactions~\cite{Klu09}.

The relation between $m_{v,0}^{\ast}$ or $\kappa$ and $E_x$ can be understood from the approximate relation~\cite{Lip89, Tri08}
\begin{equation}
E_x=\sqrt{\frac{m_1}{m_{-1}}}.
\end{equation}
While the inverse energy weighted sum rule $m_{-1}$ [Eq.~(\ref{Eq:mk})] is related to the electric dipole polarizability via Eq.~(\ref{Eq:AlphaDSm1}), the energy weighted sum rule $m_1$ depends inversely on the isovector effective mass $m_{v,0}^{\ast}$ at saturation density in the conventional Skyrme mean-field model~\cite{Zha15a}. The latter can be understood by the fact that the $m_1$ can be expressed as~\cite{Col13}
\begin{equation}
m_1=\frac{9}{4\pi}\frac{1}{2m}\frac{NZ}{A}(1+\kappa),
\end{equation}
where the factor $\kappa$ is inversely proportional to $m_{v,0}^{\ast}$.

For $^{68}$Ni, panel (b) of Fig.~\ref{Fig:Sgdr} shows that there is a so-called  pygmy dipole resonance  around 9.55 MeV in the experimental strength function that carries 2.8(5)\%  of the energy weighted sum rule $m_1$~\cite{Ros13}. Due to the smearing of the RPA states in the calculation, the pygmy resonance is not seen and appears instead as a low-energy tail in the theoretical strength function.  The RPA strength in the energy region of 0-11 MeV carries 2.7\% of the total energy weighted sum rule and is similar to that carried by the pygmy dipole resonance.  Its contribution to the dipole polarizability is 0.38 fm$^{3}$ and thus underestimates the experimental value of 1.13 fm$^{3}$~\cite{Ros13,Roc15} as for most Skyrme-type interactions.

\begin{figure}[t]
\centering
\includegraphics[width=8.cm]{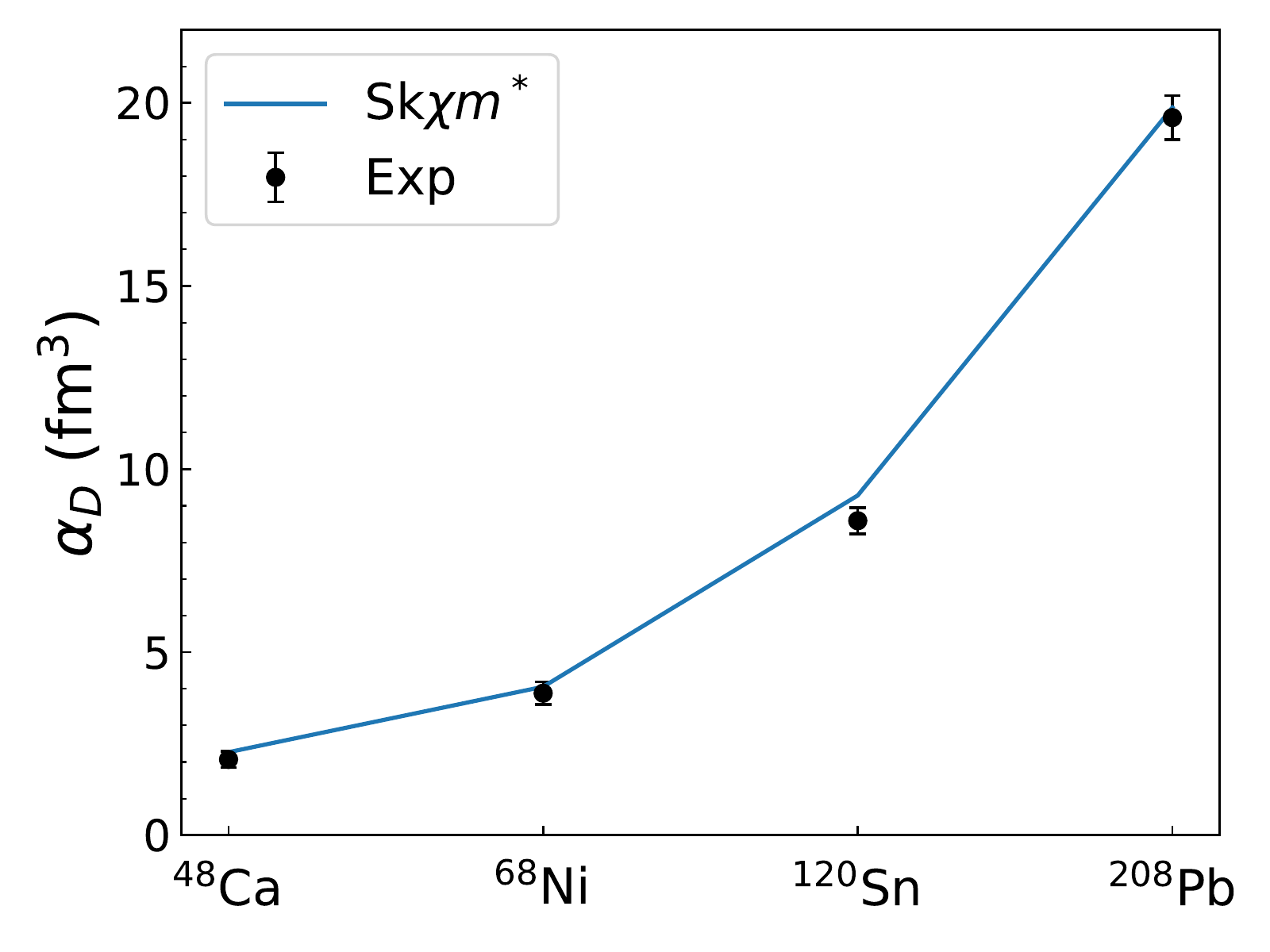}
\caption{(Color online.) Electric dipole polarizabilities of $^{48}\mathrm{Ca}$, $^{68}\mathrm{Ni}$,
$^{120}\mathrm{Sn}$, and $^{208}\mathrm{Pb}$ predicted by Sk$\chi m^*$.  For comparison, experimental data~\cite{Tam11,Ros13,Has15,Bir16} are shown as black solid circles. }\label{Fig:AlphaD}
\end{figure}

Fig.~\ref{Fig:AlphaD} exhibits the electric dipole polarizabilities of $^{48}\mathrm{Ca}$, $^{68}\mathrm{Ni}$, $^{120}\mathrm{Sn}$, and $^{208}\mathrm{Pb}$ predicted by the Sk$\chi m^{\ast}$ mean-field model together with the experimental results~\cite{Tam11,Ros13,Has15,Bir16} (detailed numerical values are listed in Table~\ref{Tab:Data}).  We find that RPA calculations using the Sk$\chi m^{*}$ interaction can overall well reproduce the experimental data. For $^{120}$Sn the prediction of Sk$\chi m^{*}$ is slightly larger than the experimental value. This could be due to the fact that, in the present work, we do not include pairing correlations, which reduce the electric dipole polarizability in the Sn isotope chain~\cite{Roc15}. In the case of $^{48}\mathrm{Ca}$, we observe that the results are consistent with recent ab initio calculations \cite{Hag16} of the dipole polarizability from chiral effective field theory. We would like to point out that we have calculated the electric dipole polarizability by using discrete RPA peaks because the $1p-1h$ RPA cannot give rise to the spreading width of nuclear giant resonances observed in experiments.  Including the spreading width, which can be done by taking into account the coupling of $1p-1h$ states with more complicated multi-particle--multi-hole configurations, is expected to reduce the calculated $\alpha_{D}$ slightly. This can be seen by smearing the RPA peaks using the Lorentzian function with a certain width. In the case of only one Lorentzian function of width $\Gamma$, the electric dipole polarizability is
at most reduced by~\cite{Roc15}
\begin{equation}
\Delta\alpha_{\mathrm{D}}\sim  \alpha_{\mathrm{D}}\frac{\Gamma^2}{4E_x^2}.
\end{equation}
According to this formula, using the experimental value of the peak energy (13.43 MeV) and width (4.07 MeV) of the IVGDR~{\cite{Die88}}, we estimate  that the reduction should be less than about $0.45~\mathrm{fm^3}$ $(\sim2\%)$ for $^{208}$Pb. Such a small correction does not influence our conclusion.

The neutron skin thickness, which has been demonstrated to be strongly correlated with
the electric dipole polarizability, is an important probe of the density slope of the nuclear symmetry energy as well as the EOS of pure neutron matter. Therefore, we also employ the Sk$\chi m^{\ast}$ mean-field model to predict the neutron skin thicknesses of $^{48}$Ca and $^{208}$Pb, and the obtained results are given by
\begin{eqnarray}
\nonumber \Delta r_{np}(\rm ^{48}Ca):  &  0.167 \pm 0.004~\mathrm{fm},\\
\nonumber \Delta r_{np}(\rm ^{208}Pb): &  0.170 \pm 0.005~\mathrm{fm}.
\end{eqnarray}
We find that the predicted results for $^{208}$Pb are in very good agreement with the constraint ${\Delta}r_{np}=0.15\pm0.03~({\text{stat.}})^{+0.01}_{-0.03}~({\text{sys.}})$\,fm
extracted from coherent pion photoproduction cross sections~\cite{Tar14} and are also consistent with the
constraint $\Delta r_{np}=0.302\pm 0.175~(\mathrm{exp}) \pm{0.026} ~(\mathrm{model}) \pm
0.005~(\mathrm{strange})$\,fm~\cite{Hor12} extracted from the parity-violating asymmetry measurement
in the Lead Radius Experiment (PREX)~\cite{Abr12}.

\section{Summary}

We have constructed a new Skyrme interaction Sk$\chi$m$^*$ by fitting the EOSs and nucleon effective masses in asymmetric nuclear matter predicted by chiral effective field theory together with the binding energies of selected closed shell doubly-magic nuclei. The new model was employed to study the isovector dipole response of $^{48}$Ca, $^{68}$Ni, $^{120}$Sn and $^{208}$Pb.  We have found that the new interaction can well reproduce the experimental data on the peak energy of the giant dipole resonance and the electric dipole polarizability. We have further calculated the neutron skin thicknesses of $^{48}$Ca and $^{208}$Pb.  The predicted neutron skin thickness of $^{208}$Pb from our study is also consistent with the experimental values~\cite{Tar14,Hor12,Abr12}. Our results thus confirm the usefulness of the new Skyrme interaction Sk$\chi$m$^*$ in studying the isovector properties of nuclei in regimes where ab initio calculations with chiral nuclear forces may not be feasible.

\vspace{.3in}

\subsection*{Acknowledgements}
This work was supported by the US Department of Energy under Contract No.\ DE-SC0015266, the National Science Foundation under Grant No.\ PHY1652199, and the Welch Foundation under Grant No.\ A-1358.

\end{document}